\documentclass[12pt]{article}

\usepackage{latexsym}
\usepackage{amsthm}
\usepackage{amsmath}
\usepackage[dvips]{graphicx}
\usepackage{amssymb}

\newcommand{\be}{\begin{equation}}
\newcommand{\ee}{\end{equation}}

\begin{document}

\baselineskip=18pt

\setcounter{footnote}{0}
\setcounter{figure}{0}
\setcounter{table}{0}

\begin{titlepage}

\begin{center}

{\Large \bf The Grassmannian Origin Of Dual \\ Superconformal Invariance}

\vspace{0.2cm}

{\bf Nima Arkani-Hamed$^a$, Freddy Cachazo$^b$ and Clifford Cheung$^{c,d}$}

\vspace{.3cm}

{\it $^{a}$ School of Natural Sciences, Institute for Advanced Study, Princeton, NJ 08540}

{\it $^{b}$ Perimeter Institute for Theoretical Physics, Waterloo, Ontario N2J W29, Canada}

{\it $^{c}$ Berkeley Center for Theoretical Physics, UC Berkeley,  Berkeley, CA 94720 \\ and \\ \it $^{d}$ Theoretical Physics Group, Lawrence Berkeley National Laboratory,  Berkeley, CA 94720}

\end{center}

\vskip.3in

\begin{abstract}
A dual formulation of the S Matrix for ${\cal N} = 4$ SYM has recently been presented, where all leading singularities of $n$-particle N$^{k-2}$MHV amplitudes are given as an integral over the Grassmannian $G(k,n)$, with cyclic symmetry, parity and superconformal invariance manifest. In this short note we show that the dual superconformal invariance of this object is also manifest. The geometry naturally suggests a partial integration and simple change of variable to an integral over $G(k-2,n)$. This change of variable precisely corresponds to the mapping between usual momentum variables and the ``momentum twistors" introduced by Hodges, and yields an elementary derivation of the momentum-twistor space formula very recently presented by Mason and Skinner, which is manifestly dual superconformal invariant. Thus the $G(k,n)$ Grassmannian formulation allows a direct understanding of {\it all} the important symmetries of ${\cal N}=4$ SYM scattering amplitudes.
\end{abstract}

\bigskip
\bigskip

\end{titlepage}

Recently a simple formula has been conjectured for all leading singularities of ${\cal N} = 4$ SYM amplitudes \cite{Gnk}. Working in twistor space, the leading singularities of single trace, color stripped, $n$-particle N$^{k-2}$MHV amplitudes are associated with the object
\begin{equation}
{\cal L}_{n;k} = \frac{1}{{\rm vol (GL}(k))} \int \frac{d^{k \times n}C_{\alpha a}}{(C_1 C_2 \cdots C_k) \cdots (C_n C_1 \cdots C_{k-1})} \prod_{\alpha = 1}^k \delta^{4|4}(C_{\alpha a} {\cal W}_a)
\end{equation}
where
\be
(C_{m_1} \cdots C_{m_k}) = \epsilon^{\alpha_1 \cdots \alpha_k} C_{\alpha_1 m_1} \cdots C_{\alpha_k m_k}
\ee
are minors of the $k \times n$ matrix $C$. Here $a=1,\cdots,n$ labels the $n$ external particles.
The integrand has a GL$(k)$ ``gauge symmetry" under which
$C_{\alpha a} \to L_{\alpha}^\beta C_{\beta a}$ for any $k \times k$ matrix $L$, and so we have to ``gauge-fix" by dividing by vol(GL$(k)$).

Going back to momentum space, the integral turns into a multi-dimensional contour integral. In \cite{Gnk}, substantial evidence was given that the residues of the integrand provide a basis for constructing tree amplitudes as well as 1- and 2-loop leading singularities. Multi-dimensional residue theorems give rise to a large number of remarkable identities between the residues, which guarantee the equivalence of many different representations of the same amplitude and enforce the cancelation of non-local poles as well as consistent infrared structure at loop level.
In \cite{Gnk}, it was also shown that the action of cyclic, parity and superconformal symmetries on eqn.~(1) is manifest. In this short note we show that the dual superconformal invariance \cite{DCI0,DCI1,DCI2} of eqn.~(1) is also manifest.

Let us immediately transform eqn.~(1) back into momentum space
\be
{\cal L}_{n;k} \! =\! \frac{1}{{\rm vol (GL}(k))}
\!\int\!\! \frac{d^{k \times n}C_{\alpha a}}{
(C_1 C_2 \cdots C_k) \cdots (C_n \cdots C_{k-1})}\! \prod_\alpha \!\delta^4(C_{\alpha a} \tilde \eta_a)
\delta^2(C_{\alpha a} \tilde \lambda_a)\!\! \int\!\! d^2\! \rho_\alpha \delta^2(\lambda_a \! - C_{\alpha a} \rho_\alpha)
\ee

In this form the geometric character of the integral over $C$, already discussed in \cite{Gnk}, is completely clear. The $\lambda_a,\tilde\lambda_a$ are 2-planes in the $n$-dimensional space, while $C$ is a $k$-plane. The integral is then over the space of $k$-planes in $n$-dimensions--the Grassmannian $G(k,n)$. The first set of bosonic delta functions forces $C$ to be orthogonal to the $\tilde \lambda$ plane, while the second forces some linear combination of the $k$ $n-$vectors in $C$ to equal $\lambda$; in other words, the $k$-plane $C$ must also contain the 2-plane $\lambda$. Note that these two requirements can be satisfied only if the $\lambda$ plane is orthogonal to the $\tilde \lambda$ plane -- $\sum \lambda_a \tilde \lambda_a = 0$ -- which is nothing other than the statement of momentum conservation\footnote{Note that in writing eqn.~(1) we are working with real variables in (2,2) signature, and in the $n$-dimensional space there is a natural notion of ``orthogonality" associated with the standard dot product. As discussed in \cite{Gnk}, the final, fully gauge-fixed form of the expression for ${\cal L}_{n;k}$ is defined for arbitrary complex momenta.}.

Given that the $C$ plane must contain the $\lambda$ plane, it is very natural to split $C$ into a part that literally {\it is} the $\lambda$ 2-plane, and only perform the integral over the remaining $(k-2)$ directions. This motivates us to try and write the resulting expression as an integral over
$G(k-2,n)$. This is essentially trivial -- the only (minor) complication is that the minors appearing in eqn.~(1) are $k \times k$ minors, not $(k-2) \times (k-2)$ minors. But from high-school algebra we are familiar with a natural linear transformation which maps the larger minors to the smaller minors. As we will see this linear transformation takes us from the momentum variables to the ``momentum twistor" variables recently introduced in a remarkable paper by Hodges \cite{MT}. The remaining integral over $G(k-2,n)$ precisely yields the momentum-twistor Grassmannian formula very recently presented by Mason and Skinner \cite{MS}. This expression is manifestly dual superconformal invariant. In this way momentum twistors and dual superconformal invariance are seen to follow very naturally from the Grassmannian formulation, and we see that eqn.~(1) provides a simple understanding of {\it all} the important symmetries of the theory.

To get started, let us turn to ``gauge-fixing" the GL$(k)$ symmetry, which we can do by introducing a gauge-fixing
function $\Delta_{{\rm GL(k)}}$ which can in general be a function of $C_{\alpha a}$ and $\rho_\alpha$, and writing
\be
{\cal L}_{n;k} = \int \frac{d^{k \times n}C_{\alpha a} ~ \Delta_{{\rm GL(k)}}}{(C_1 C_2 \cdots C_k) \cdots (C_n C_1 \cdots C_{k-1})} \prod_\alpha \delta^4(C_{\alpha a} \tilde \eta_a)
\delta^2(C_{\alpha a} \tilde \lambda_a) \int d^2 \rho_\alpha \delta^2(\lambda_a - C_{\alpha a} \rho_\alpha)
\ee
In practice, we can choose $\Delta_{{\rm GL}(k)}$ to be a product of $k^2$
delta functions fixing some components of $C,\rho$ to some canonical values. For instance in \cite{Gnk}, we implicitly chose $\Delta$ to only depend on $C$, and used it to fix $k$ of the columns of $C$ to an orthonormal basis. This gauge-fixing was convenient for showing the parity invariance of ${\cal L}_{n;k}$, as well as making the connection between eqn.~(1) and the ``link representation" of \cite{link}. In this note we will find another gauge-fixing convenient, the analog of which was already used in \cite{Gnk} in the discussion of MHV amplitudes.

Since our ultimate goal is to be left with an integral over $k-2$ planes, we wish to start by only partially gauge-fixing, leaving a GL$(k-2)$ to be fixed only at the end. Therefore, we start by imposing $k^2 - (k-2)^2 = 4k - 4 = 2k + 2 (k-2)$ conditions. In other words we will write
\be
\Delta_{GL(k)} = \Delta_{GL(k) \to GL(k-2) \times T_2} \times \Delta_{GL(k-2) \times T_2 \to GL(k-2)} \times \Delta_{GL(k-2)}
\ee
which gauge fix the GL$(k)$ in steps, first down to GL$(k-2) \times T_2$ where $T_2$ are translations by a vector in the $\lambda$ plane, then
to GL$(k-2)$. In the first step, we use $2k$ delta functions to gauge fix the two $k$-vectors $\rho$ to
\be
\rho = \left(\begin{array}{cc} 1 & 0 \\ 0 & 1 \\ 0 & 0 \\ \vdots & \vdots \\ 0 & 0 \end{array} \right)
\ee
Formally we choose
\be
\Delta_{GL(k) \to GL(k-2) \times T_2} = \prod_\alpha \delta^2(\rho_{\overline{\alpha} \alpha} - \delta_{\overline{\alpha} \alpha}).
\ee
Here we use $\overline{\alpha} = 1,2$ to label the spinor Lorentz indices. These equation fix all the $\rho$ variables; the $\delta^2(\lambda_a - C_{\alpha a} \rho_\alpha)$ in the integral then fixes the top two rows of $C$ to be the $\lambda$ 2-plane, that is $C_{{\overline \alpha} a} = \lambda_{\overline{\alpha} a}$, or
\be
C = \left(\begin{array}{cccc} \lambda_{\overline{\alpha} = 1, 1} & \lambda_{\overline{\alpha} = 1, 2} & \cdots & \lambda_{\overline{\alpha} = 1, n} \\ \lambda_{\overline{\alpha} = 2, 1} & \lambda_{\overline{\alpha} = 2, 2} & \cdots & \lambda_{\overline{\alpha} = 2, n} \\ C_{3,1} & C_{3,2} & \cdots & C_{3,n} \\ \vdots & \vdots & \ddots & \vdots \\ C_{k,1} & C_{k,2} & \cdots & C_{k,n} \end{array} \right)
\ee
With this form we can factor out the momentum and super-momentum conserving delta functions trivially, since
\be
\prod_{\alpha=1}^k \delta^4(C_{\alpha a} \tilde \eta_a) \delta^2(C_{\alpha a} \tilde \lambda_a) = \delta^4(\lambda_a \tilde \lambda_a) \delta^8(\lambda_a \tilde \eta_a) \times \prod_{\hat{\alpha}=3}^k \delta^4(C_{\hat{\alpha} a} \tilde \eta_a) \delta^2(C_{\hat{\alpha} a} \tilde \lambda_a)
\ee
where we use the index $\hat{\alpha}$ to run over the $(k-2)$ values $3, 4, \cdots, k$.

Having done this,  the unfixed part of GL$(k)$ then consists of GL$(k-2)$ transformations, together with translations $T_2$ of the remaining $(k-2)$ $n$-vectors by any vector in the $\lambda$ plane. Geometrically, the most natural way to gauge fix $T_2$ is to impose the $2 \times (k-2)$ constraints that the remaining $(k-2)$ vectors in $C$ are orthogonal to the $\lambda$ plane:
\be
\label{t2fix}
\Delta_{GL(k-2) \times T_2 \to GL(k-2)} = J \prod_{\hat{\alpha}} \delta^2(C_{\hat{\alpha} a} \lambda_a)
\ee
Here $J$ is some Jacobian that only depends on the $\lambda$'s. Here we have used the index $\hat{\alpha} = 3,4, \cdots ,k$ to denote the remaining $(k-2)$ rows of $C$.

We have thus arrived at
\be
\label{almost}
{\cal L}_{n;k} =  \frac{J \delta^4(\lambda_a \tilde \lambda_a) \delta^8(\lambda_a \tilde \eta_a)}{{\rm vol}({\rm GL}(k-2))} \int \frac{d^{(k-2) \times n} C_{\hat{\alpha} a}}{(C_1 C_2 \cdots C_k) \cdots (C_n C_1 \cdots C_{k-1})} \prod_{\hat{\alpha}} \delta^2(C_{\hat{\alpha} a} \lambda_a) \delta^2(C_{\hat{\alpha} a} \tilde \lambda_a) \delta^4(C_{\hat{\alpha} a} \tilde \eta_a)
\ee

This form is almost what we want: an integral manifestly over a space of $(k-2)$ planes in $n$ dimensions\footnote{Note that for general $2k > n$, $G(k-2,n)$ is not contained in $G(k,n)$. Geometrically, the space of $k$-planes that are forced to contain a given $2$-plane is the same of the space of $(k-2)$ planes living in $(n-2)$ dimensions: $G(k-2,n-2)$ {\it is} contained in $G(k,n)$. The $\delta^2(C \cdot \lambda)$ factors in eqns.~(\ref{t2fix},\ref{almost}) are reducing the integral over $G(k-2,n)$ to one over $G(k-2,n-2)$, but in a moment we will find these delta functions nicely unify with the remaining ones, and the resulting object is naturally integrated over $G(k-2,n)$.}. However the problem is that the minors appearing in the denominator of the measure involve the full $k \times n$ matrix $C$, whose first two rows are the $\lambda$ plane. It is natural to try and find a linear transformation that maps the $k \times k$ minors to $(k-2) \times (k-2)$ minors. Consider the minor $(C_1 C_2\cdots C_{k-1}C_k)$. We will take linear transformations of the columns of $C$, so that the first two rows of the middle $2,\cdots,k-1$ columns are set to zero. The simplest way to do this is to define new $k$ vectors $D_{\alpha b}$ as
\be
D_{\alpha b} = x_b C_{\alpha b-1} + y_b C_{\alpha b} + z_b C_{\alpha b+1}
\ee
Clearly $D_2,\cdots,D_{k-1}$ are a linear combination of $C_1, \cdots, C_k$, therefore
\be
(C_1 D_2 \cdots D_{k-1} C_k) = f(x,y,z) (C_1 \cdots C_k)
\ee
We choose $x,y,z$ so that $D_{\overline{\alpha} b} = 0$. This fixes $D$ to be of the form
\be
D_{\alpha b} = q_b \left(\langle b+1 \, b \rangle C_{\alpha b-1} + \langle b-1 \, b+1 \rangle C_{\alpha b} + \langle b \, b-1 \rangle C_{\alpha b+1}\right)
\ee
where $q_b$ is any normalization factor. Here $D_{\overline \alpha b} = 0$ due to the Schouten identity $\langle b+1 \, b \rangle \lambda_{b-1} + \langle b-1 \, b+1 \rangle \lambda_b + \langle b \, b-1 \rangle \lambda_{b+1} = 0$. Clearly
\be
(C_1 D_2 \cdots D_{k-1} C_k) = \langle \lambda_1 \lambda_k \rangle (D_2 \cdots D_{k-1})
\ee
where $(D_2 \cdots D_{k-1})$ is the $(k-2) \times (k-2)$ minor made from the non-zero elements $D_{\hat{\alpha} b}$:
\be
(D_2 \cdots D_{k-1}) = \epsilon^{\hat{\alpha}_1 \cdots \hat{\alpha}_{k-2}} D_{\hat{\alpha}_1 2} \cdots D_{\hat{\alpha}_{k-2} k-1}
\ee
Thus we have succeeded in mapping the $k \times k$ minors to $(k-2) \times (k-2)$ minors up to a $\lambda$ dependent factor
\be
\label{minormap}
(C_1 \cdots C_k) = J^\prime (D_2 \cdots D_{k-1})
\ee

What we have done is to introduce an $n \times n$ linear transformation $Q_{ab}$
\be
D_{\alpha b} = C_{\alpha a} Q_{ab}
\ee
where
\be
Q_{ab} = q_b \times \left(\langle b - 1 \, b+1 \rangle \delta_{ab} + \langle b \, b-1\rangle \delta_{a,b+1} + \langle b+1 \, b\rangle \delta_{a,b-1} \right)
\ee
satisfying
\be
\lambda_a Q_{ab} = 0
\ee
Now, using $Q$, we can re-write
\be
\label{andrew}
\tilde \lambda_a = Q_{ab} \mu_b, \, \, \tilde \eta_a = Q_{ab} \eta_b
\ee
expressing $\tilde \lambda_a, \tilde \eta_a$ in terms of new variables $\mu_b,\eta_b$; these forms guarantee that both momentum and super-momentum are conserved, since e.g. $\lambda_a \tilde \lambda_a = \lambda_a Q_{ab} \mu_b = 0$. There is a natural choice for the normalization $q_b$ so that the
$\mu_b,\eta_b$ have nice little-group transformation properties under $\lambda_b \to t_b \lambda_b, \tilde \lambda_b \to t_b^{1} \tilde \lambda_b, \tilde \eta_b \to t_b^{-1} \tilde \eta_b$. Choosing $q_b = \frac{1}{\langle b-1 \, b \rangle \langle b \, b+1 \rangle}$, $(\mu_b,\eta_b) \to t_b (\mu_b,\eta_b)$ have exactly the same little group transformation properties as $\lambda_b$. Thus we choose
\be
\label{finalQ}
Q_{ab} = \frac{\langle b - 1 \,  b+1 \rangle \delta_{ab} + \langle b \, b-1 \rangle \delta_{a,b+1} + \langle b+1 \, b \rangle \delta_{a,b-1}}{
\langle b+1 \, b \rangle \langle b \, b-1 \rangle}
\ee
With this choice for $Q$ the relationship between the $D$ minors and $C$ minors becomes
\be
(D_2 \cdots D_{k-1}) = \frac{1}{\langle 1 2 \rangle \langle 2 3 \rangle \cdots \langle k-1 \, k \rangle} (C_1 \cdots C_k)
\ee
where we have given the $\lambda$ dependent factor explicitly.
Note also that with this choice
\be
Q_{ab} = Q_{ba}
\ee

Now, we would clearly like to change variables from $(k-2) \times n$ variables $C_{\hat{\alpha} a}$ to another set of $(k-2) \times n$ variables $D_{\hat{\alpha} b} = C_{\hat{\alpha} a} Q_{ab}$.  Not only are the $k \times k$ minors of $C$ equal (up to $\lambda$ dependent factors) to $(k-2) \times (k-2)$ minors of $D$, but also the $\delta$ functions involving $\tilde \lambda$ and $\tilde \eta$ are directly functions of $D$, since
\be
\delta^2(C_{\hat{\alpha} a} \tilde \lambda_a) = \delta^2(D_{\hat{\alpha} b} \mu_b), \, \, \delta^4(C_{\hat{\alpha} a} \tilde \eta_a) = \delta^4(D_{\hat{\alpha} b} \eta_b)
\ee
But we can't literally make this change of variables, since $Q_{ab}$ is not invertible, given that $\lambda_a Q_{ab} = 0$.  Any two $C_{\hat{\alpha} a}$, which differ by any translation in the $\lambda$ 2-plane, will yield the same $D$. However, this translational freedom $T_2$ was precisely what we gauge-fixed by including the delta function $\prod_{\hat{\alpha}} \delta^2(C_{\hat{\alpha} a} \lambda_a)$. Thus, given some fixed $D_{\hat{\alpha} a}$, if there is any solution of $D_{\hat{\alpha} b} = C_{\hat{\alpha} a} Q_{ab}$ satisfying the $ 2\times (k-2)$ constraints $C_{{\hat \alpha} a} \lambda_a = 0$, this solution is unique. On the other hand to be able to find ${\it any}$ solution
$C_{\hat{\alpha} a}$ to $D_{\hat{\alpha} b} = C_{\hat{\alpha} a} Q_{ab}$, $D$ can not be a completely general $(k-2) \times n$ matrix, but must satisfy $2 \times (k-2)$ constraints. Since $Q_{ab} = Q_{ba}$ and so $Q_{ab} \lambda_b = 0$, these constraints are clearly $D_{\hat{\alpha} b} \lambda_b = C_{\hat{\alpha} a} Q_{ab} \lambda_b = 0$. We therefore conclude that
\be
\label{ctod}
\int d^{(k-2) \times n} C_{\hat{\alpha} a} \prod_{\hat{\alpha}} \delta^2(C_{\hat{\alpha} a} \lambda_a) f(C_{\alpha a} Q_{ab})= J^{ \prime \prime} \int d^{(k-2) \times n} D_{\hat{\alpha} b} \prod_{\hat{\alpha}} \delta^2(D_{\hat{\alpha} b} \lambda_b) f(D_{\hat{\alpha} b})
\ee
here $J^{\prime \prime}$ is yet another Jacobian that only depends on the $\lambda$'s. We can arrive at this result more formally by beginning with the LHS of eqn.~(\ref{ctod}), and multiply by $1$ in the form
\be
\label{ddelta}
1 = \int d^{(k-2) \times n} D_{\hat{\alpha} b} \prod_{\hat{\alpha}, b} \delta(D_{\hat{\alpha} b} - C_{\hat{\alpha} a} Q_{ab})
\ee
We can then perform the integrals over $C$. The $\delta^2(C \cdot \lambda)$ factors together with all but $2 \times (k-2)$ of the delta functions introduced in eqn~(\ref{ddelta}) allow us to uniquely solve for all the $C$'s, leaving us with a remaining $2 \times (k-2) \, \delta$ functions linear in the $D$'s, which we know must impose $2 \times (k-2)$ constraints $D \cdot \lambda = 0$. This yields the RHS of eqn.~(\ref{ctod}) up to a $\lambda$ dependent Jacobian factor $J^{\prime}$.

We are now done: using eqns.~(\ref{almost},\ref{minormap}, \ref{ctod}) we find
\be
{\cal L}_{n;k} = J^{\prime \prime \prime} \delta^4(\lambda_a \tilde \lambda_a) \delta^8(\lambda_a \tilde \eta_a)
\frac{1}{{\rm vol}({\rm GL}(k-2))}\! \int \!\!\frac{d^{(k-2) \times n} D_{\hat{\alpha} b}}{(D_1 D_2 \cdots D_{k-2}) \cdots (D_n D_1 \cdots D_{k-3})} \delta^{4|4}(D_{\hat{\alpha} b}{\cal Z}_b^D)
\ee
where $J^{\prime \prime \prime}$ is the product of all the various $\lambda$-dependent prefactors and Jacobians above, and ${\cal Z}_b^D$ are precisely the ``momentum twistors" defined by Hodges \cite{MT}
\be
{\cal Z}^D = \left(\begin{array}{c} \lambda \\ \mu \\ \eta \end{array}\right)
\ee
related to the original momentum space variables via eqns.~(\ref{andrew},\ref{finalQ}):
\begin{eqnarray}
\tilde \lambda_b = \frac{\langle b+1 \, b \rangle \mu_{b-1} + \langle b-1 \, b+1 \rangle \mu_{b} + \langle b \, b-1 \rangle \mu_{b+1}}{\langle b-1 \, b \rangle \langle b \, b+1 \rangle} \\
\tilde \eta_b = \frac{\langle b+1 \, b \rangle \eta_{b-1} + \langle b-1 \, b+1 \rangle \eta_{b} + \langle b \, b-1 \rangle \eta_{b+1}}{\langle b-1 \, b \rangle \langle b \, b+1 \rangle}
\end{eqnarray}
We see that the linear operator $Q_{ab}$ so naturally motivated by the geometrical considerations of reducing our initial integral over
$G(k,n)$to the natural integral over $G(k-2,n)$ precisely performs the change of variable from spinor-helicity variables to momentum twistors!

The overall $\lambda$-dependent factor $J^{ \prime \prime \prime}$ can simply be determined by explicit computation
\be
J^{\prime \prime \prime} = \frac{1}{\langle 1 2 \rangle \langle 2 3 \rangle \cdots \langle n 1 \rangle}
\ee
which nicely completes the $\delta$ function prefactors into the full MHV amplitude superamplitude:
\be
\label{MS1}
 {\cal L}_{n;k} = \frac{\delta^4(\lambda_a \tilde \lambda_a) \delta^8(\lambda_a \tilde \eta_a)}{\langle 1 2 \rangle \cdots \langle n 1 \rangle} \times {\cal R}_{n;k}
\ee
with
\be
\label{MS2}
 {\cal R}_{n;k} = \frac{1}{{\rm vol}({\rm GL}(k-2))} \int \frac{d^{(k-2) \times n} D_{\hat{\alpha} b}}{(D_1 D_2 \cdots D_{k-2}) \cdots (D_n D_1 \cdots D_{k-3})} \delta^{4|4}(D_{\hat{\alpha} b}{\cal Z}_b^D)
\ee
This formula for ${\cal R}_{n;k}$ is precisely the one recently presented by Mason and Skinner \cite{MS}, and makes the {\it dual} superconformal invariance of ${\cal R}_{n;k}$ completely manifest. Note that after GL$(k-2)$ gauge fixing and using the $4k$ bosonic delta functions, we are left with an integral over $(k-2)(n-k-2)$ variables, which is of course precisely the same as the momentum-space integrals studied in \cite{Gnk}. The residues studied in \cite{Gnk} are also trivially related to those of eqn.~(\ref{MS2}): setting the minor $(C_iC_{i+1}\ldots C_{i+k-1})$ to zero in \cite{Gnk} corresponds to setting the minor $(D_{i+1}D_{i+2}\cdots D_{i+k-2})$ to zero in eqn.~(\ref{MS2}). Therefore, all the results obtained in \cite{Gnk}, identifying the various contours associated with tree amplitudes as well as 1- and 2-loop leading singularities, are trivially translated to the corresponding contours for ${\cal R}_{n;k}$ as well.

It is remarkable that the original formula eqn.~(1) in the usual twistor space is almost identical in form to eqn.~(\ref{MS1}) in the dual momentum-twistor space, differing only by factoring out the MHV superamplitude and shifting $k \to k-2$. This similarity implies an infinite chain of interesting dualities: we can take the momentum-twistor formula in $G(k-2,n)$ and go to {\it its} dual space, factoring out the MHV superamplitude to get an expression in $G(k-4,n)$. In this way we can relate an amplitude with some given $k$ to one with $k \to k - 2 \to k- 4 \cdots$. This shows that in some sense, amplitudes with even (odd) $k$ can be related to MHV (NMHV) amplitudes in some dual space. Furthermore, due to parity we can also send $k \to n-k$ at any stage in this process and increase $k$ as well, yielding an infinite number of dual descriptions.

We have seen that the $G(k,n)$ integral of eqn.~(1) simultaneously manifests the superconformal and dual superconformal symmetries, as well as cyclic and parity invariance. It is amusing that different ``gauge-fixings" of the GL$(k)$ symmetry are convenient for seeing different symmetries: the gauge-fixing introduced in \cite{Gnk} is more convenient for manifesting parity, while the one exploited in this paper is better suited to manifesting dual superconformal invariance. Since the full Yangian algebra of ${\cal N} = 4$ SYM scattering amplitudes \cite{Yangian} arises from taking commutators of superconformal and dual superconformal generators, we expect that the object in eqn.~(1) is invariant under the full Yangian as well. Indeed, as mentioned in \cite{Gnk}, perhaps eqn.~(1) should be thought of as a generating function for all Yangian invariants, with residue theorems simply encoding all the remarkable relations between these invariants. It would be interesting to establish the Yangian invariance of eqn.~(1) even more directly.

\section*{Acknowledgments}

We thank Fernando Alday, Jacob Bourjaily, Jared Kaplan, Juan Maldacena, Edward Witten and especially Andrew Hodges, Lionel Mason and David Skinner for stimulating discussions. N.A.-H. is supported by the DOE under grant DE-FG02-91ER40654, F.C. was supported in part by the NSERC of Canada and MEDT of Ontario.

\end{document}